\documentclass[]{spie}  

\usepackage{aas_macros}
\usepackage{amsmath,amsfonts,amssymb}
\usepackage{graphicx}
\usepackage[colorlinks=true, allcolors=blue]{hyperref}

\title{Exploring reinforcement learning to enhance focal-plane wavefront control for vortex coronagraphs}

\author[a]{İremsu Taşkın} 
\author[b]{Jalo Nousiainen}
\author[a]{Gilles Orban de Xivry}
\author[a]{Olivier Absil} 
\author[b]{Markus Kasper}
\affil[a]{STAR Institute, Universit\'e de  Li\`ege, All\'ee du Six Ao\^ut 19c, 4000 Li\`ege, Belgium}
\affil[b]{European Southern Observatory, Garching bei München, Germany}

\authorinfo{Send correspondences to İremsu Taşkın\\ E-mail: iremsu.taskin@uliege.be}

\begin{document} 
\maketitle

\begin{abstract}
High Contrast Imaging (HCI) on ground-based telescopes suffers from phase aberrations on the observed wavefront caused by atmospheric turbulence. Adaptive Optics (AO) systems are adept at correcting these aberrations, but fall short in the correction of non-common path aberrations (NCPAs). NCPAs arise because the wavefront sensor (WFS) measures and corrects a wavefront that is different from that affecting the science images, thus requiring additional intervention. This work makes use of focal-plane wavefront sensing and reinforcement learning (RL) to address the wavefront aberrations caused by NCPAs. The PO4NCPA algorithm utilizes sequential phase diversity to address phase ambiguities and is tested on a simulation designed for the Mid-infrared ELT Imager and Spectrograph (METIS) instrument. In this paper, we present the performance of PO4NCPA with scalar and vector vortex coronagraphs to demonstrate its flexibility.
\end{abstract}


\section{INTRODUCTION}
\label{sec:intro} 

According to the NASA Exoplanet Archive \cite{NASA_exo_detection_tool,2025Christiansen_NASA_tool}, over 6000 exoplanets have been detected to date, most of the detections being through indirect methods such as transit observations and radial velocity calculations. While direct imaging technique has led to less than 100 exoplanet detections so far \cite{2023Currie}, it remains one of the most prominent methods in exoplanet studies. High contrast imaging (HCI) is a breakthrough technology that allows for direct imaging of exoplanets and provides the information necessary to understand the formation and evolution of the target \cite{2005Guyon}. This technique suppresses the stellar light in order to reveal the planet image. 

The high contrast ratio between a host star and the planet as well as the close proximity of the planet to the host star make HCI quite challenging. Ground based telescopes address these challenges by implementing Extreme Adaptive Optics (XAO) \cite{2018Guyon_XAO} that achieve high contrast in low angular separations and coronagraphs that suppress starlight. However, the system can still suffer from residual wavefront aberrations such as low wind effect (LWE), residual halo \cite{2018Guyon_XAO} and slow-evolving non-common path aberrations (NCPA). NCPAs can be caused by chromatic aberrations or aberrations caused by the optics that present themselves in the wavefront that is seen by the science camera, but is not present in the wavefront seen by the wavefront sensor (WFS). As NCPAs are caused by the optical path difference between the WFS and the science camera, focal plane wavefront sensing (FPWFS) is an attractive technique to address them. Identifying NCPAs using FPWFS experiences a problem in which even modes with opposite signs in the pupil plane can create identical images at the focal plane if there is a symmetric pupil. This phenomenon is referred to as sign ambiguity. Various methods have been used to lift the sign ambiguity in FPWFS observations. Previous studies have lifted the sign ambiguity by introducing known even phase aberrations (phase diversity \cite{1982Gonsalves}), pupil asymmetries \cite{2013Martinache, 2024Orban_asymm} or sequential phase diversity \cite{2010Gonsalves, 2012Keller, Korkiakoski:14}. 

Focal plane wavefront sensing must solve a non-linear inverse problem in order to obtain pupil plane phase aberrations from focal plane intensities. Considering this non-linear relationship and the complexity of the techniques that break the sign ambiguity, machine learning (ML) methods are increasingly preferred in FPWFS setups. Previous works have used deep supervised learning to correct NCPAs \cite{2021Orban,2022Quesnel}, but these works rely on large datasets to train the neural network. Reinforcement learning (RL) methods are model-free alternatives that can be applied to FPWFS. Our work utilizes the RL framework developed in Ref.~\citenum{2022Nousiainen} and Ref.~\citenum{2024Nousiainen_lab} to FPWFS and uses sequential phase diversity to correct NCPAs. 

This paper builds upon our previous demonstration \cite{2026Nousiainen} of PO4NCPA for FPWFS using sequential phase diversity. Ref.~\citenum{2026Nousiainen} presented the results of PO4NCPA's performance for standard imaging, perfect coronagraph and vector vortex coronagraph. Here, we present an in-depth analysis of PO4NCPA in two distinct setups: one with a scalar vortex coronagraph and one with a vector vortex coronagraph. We present the results of our numerical experiments for both static and dynamic NCPAs. As we are interested in the possible implementation of PO4NCPA on the Extremely Large Telescope (ELT) Mid-Infrared ELT Imager and Spectrograph (METIS) \cite{2024Brandl}, we simulate an ELT-like telescope. A big concern in N-band observations is the water vapor (WV) seeing, as it has the strongest effects in this band. In line with this, we focus on NCPAs that arise from the WV seeing \cite{2022Absil_wv}, which is expected to significantly affect the N-band METIS observations.

\section{METHODOLOGY}

\subsection{Focal Plane Wavefront Sensing}
\label{sec:FPWFS}

An optical telescope with diameter $\mathrm{D}$ observing a target at wavelength $\lambda$ is diffraction limited at $\lambda/\mathrm{D}$. However, this idealized resolution is rarely reached due to wavefront aberrations. Modern telescopes use deformable mirrors (DM) and wavefront sensors (WFS) to compensate for these aberrations in an attempt to get close to the diffraction limit. While such a setup can address atmospheric aberrations, its performance is limited due to NCPAs. Most WFS have optical paths different from the science detector, separated by a beamsplitter or a dichroic. Thus, the wavefronts reaching the WFS and the science detector are different. Consequently, the WFS can not address the NCPAs (comprising non-common path and chromatic aberrations) that the image at the science detector suffers from. FPWFS uses the focal plane images to calculate the pupil plane wavefront, placing the FPWFS in the same optical path as the science detector. The properties and advantages of using FPWFS for XAO systems is discussed in detail in Ref.~\citenum{2005Guyon} and Ref.~\citenum{2018Jovanovic}.

In FPWFS, the Fourier relationship between the focal plane point spread function (PSF) and the pupil plane phase aberration leads to sign ambiguity for even radial order Zernike modes. As a result, even Zernike modes produce the same image in the focal plane regardless of their sign for circularly symmetric pupils. This phenomenon is further discussed in Ref.~\citenum{2006Mugnier}. The sign ambiguity introduced for even Zernike modes is
\begin{equation}
\label{eq:sign_amb}
|\mathcal{F}(E(x, y))|^2 =  |\mathcal{F}(E^{*}(-x, -y))|^2,
\end{equation}
where $E(x,y) = \exp\!\left(-i\phi(x,y)\right)$ is the pupil-plane electric field with phase aberrations $\phi$ and $E^*$ is the conjugate. For a centro-symmetric pupil and even modes, this relation translates to an indetermination of the sign on the even part of the phase.

Historically, various methods have been used to address sign ambiguities in FPWFS. In one of the earliest implementations of FPWFS, Ref.~\citenum{1982Gonsalves} used a second image with a known wavefront aberration to break the ambiguity and recover the first image. More recent research has used this technique with various predetermined aberrations such as a defocused image \cite{2022Quesnel}. Alternatively, past focal plane observations and corrections of NCPAs can be used to lift the sign ambiguity. This technique is referred to as sequential phase diversity and is described initially in Ref.~\citenum{2010Gonsalves}. In this work, we make use of sequential phase diversity without any predefined aberrations. A similar approach has been implemented in the Fast and Furious \cite{Korkiakoski:14} algorithm and validated on sky \cite{2020Bos}, but it is not compatible with coronagraphs. Further work in Ref.~\citenum{2023Bottom} has been able to address this limitation only for symmetric coronagraphs. Another alternative that is compatible with all coronagraphs was presented in the same paper, but it uses a deep learning (DL) approach that requires large datasets \cite{2023Bottom}. Here, we use RL in order to address sign ambiguities for FPWFS applicable for all coronagraph types.

\subsection{Vortex Coronagraphs}
\label{sec:vortex}

A vortex coronagraph (VC) \cite{2005Mawet_vortex} is a vortex focal plane mask coupled with a downstream pupil plane Lyot stop where the phase mask diffracts the on-axis light outside of the pupil and the Lyot stop blocks the diffracted light. The vortex phase mask creates a phase screw $\exp{(il_{p}\phi)}$ where $l_p$ is the topological charge and $\phi$ is the azimuthal coordinate \cite{2012Riaud, 2024Orban_vortex}. This coronagraph allows for high contrast imaging of exoplanets where the on-axis starlight is suppressed while the light from off-axis exoplanet remains within the pupil. Theoretically, the VC provides a perfect cancellation of the stellar light in the absence of any wavefront error and for a circular pupil. As for most coronagraphs, the  VC is sensitive to low-order aberrations which directly affect its rejection performance\cite{2010Mawet, 2016SPIE_vortex}.

There are two subtypes of vortex coronagraphs, vector vortex coronagraph (VVC) and scalar vortex coronagraph (SVC). The VVC applies two opposite phase screws \cite{2012Riaud}, thus applying conjugated phase ramps to each circular polarization state \cite{2024Orban_vortex}. This is denoted by the topological charge $\pm|l_p|$. In this work, we use $|l_p| = 2$ as it is the most commonly used topological charge \cite{2022Quesnel}. Conversely, the SVC applies the same phase ramp to each circular polarization state as it is based on longitudinal phase delays. This behaviour is similar to the behaviour that the VVC would have on a single circular polarization. 

Similarly to non-coronagraphic imaging, for the VVC, two wavefronts with opposite signs for even Zernike modes produce the same image after propagation, as illustrated in Ref.~\citenum{2022Quesnel}. While the VVC shows ambiguity to all even modes, the SVC experiences ambiguity only to pure radial even modes ($m=0$, corresponding to the first mode of each even radial order when denoted by Noll's sequential indices \cite{1976Noll_zernike}). A detailed analysis of the phase retrieval properties of the VVC and SVC to even modes can be found in Ref.~\citenum{2024Orban_vortex}.

\subsection{Reinforcement Learning}
\label{sec:RL}

As we aim for an algorithm that can learn in real time and is independent of large datasets, we use RL in our work. RL algorithms train an agent and make decisions to attain a predetermined result based on data from a dynamic environment. A common framework used to formalize RL is Markov Decision Process (MDP). MDP formalizes the sequential decision-making process that relates an agent with its environment. RL employs MDP in episodes of discrete time sequences. At each timestep $t$, the environment occupies a state $\boldsymbol{s}_t$ that belongs to a set of all possible states denoted as $\mathcal{S}$. The agent, which is the decision-maker in the algorithm, selects an action $\boldsymbol{a}_t$ from the action space $\mathcal{A}$ based on the state it received from the environment. This action is sent back to the environment, which moves to a new state $\boldsymbol{s}_{t+1}$ as a response. The probability of transitioning to a new state after an action is defined by a conditional probability distribution $p(\boldsymbol{s}_{t+1}|\boldsymbol{s}_t,\boldsymbol{a}_t)$, which represents the transition dynamics. These transition dynamics are assumed to contain various unknowns such as NCPAs, optical element imperfections and LWE. At each step, a reward $r(\boldsymbol{s}_t,\boldsymbol{a}_t)$ is observed. This reward is designed so the agent makes decisions that will result in a particular outcome in the environment. 

A "policy", $\pi:\boldsymbol{s}_t \mapsto \boldsymbol{a}_t$, maps states into actions and governs the decisions made by the agent. An RL algorithm aims to optimize the policy such that the maximum cumulative reward can be achieved. In model-based RL, the true dynamics model $p(\boldsymbol{s}_{t+1}|\boldsymbol{s}_t,\boldsymbol{a}_t)$ is estimated using an approximate model $\hat{p}(\boldsymbol{s}_{t+1}|\boldsymbol{s}_t,\boldsymbol{a}_t)$ based on data. Model-free RL, on the other hand, does not attempt to learn the dynamics model at all. The application of RL algorithms for AO is discussed in Ref.~\citenum{2021Nousiainen} and FPWFS as a MDP for RL algorithms is detailed in Ref.~\citenum{2026Nousiainen}.

\begin{figure} [ht]
   \begin{center}
   \begin{tabular}{l | r} 
   \includegraphics[height=4.5cm]{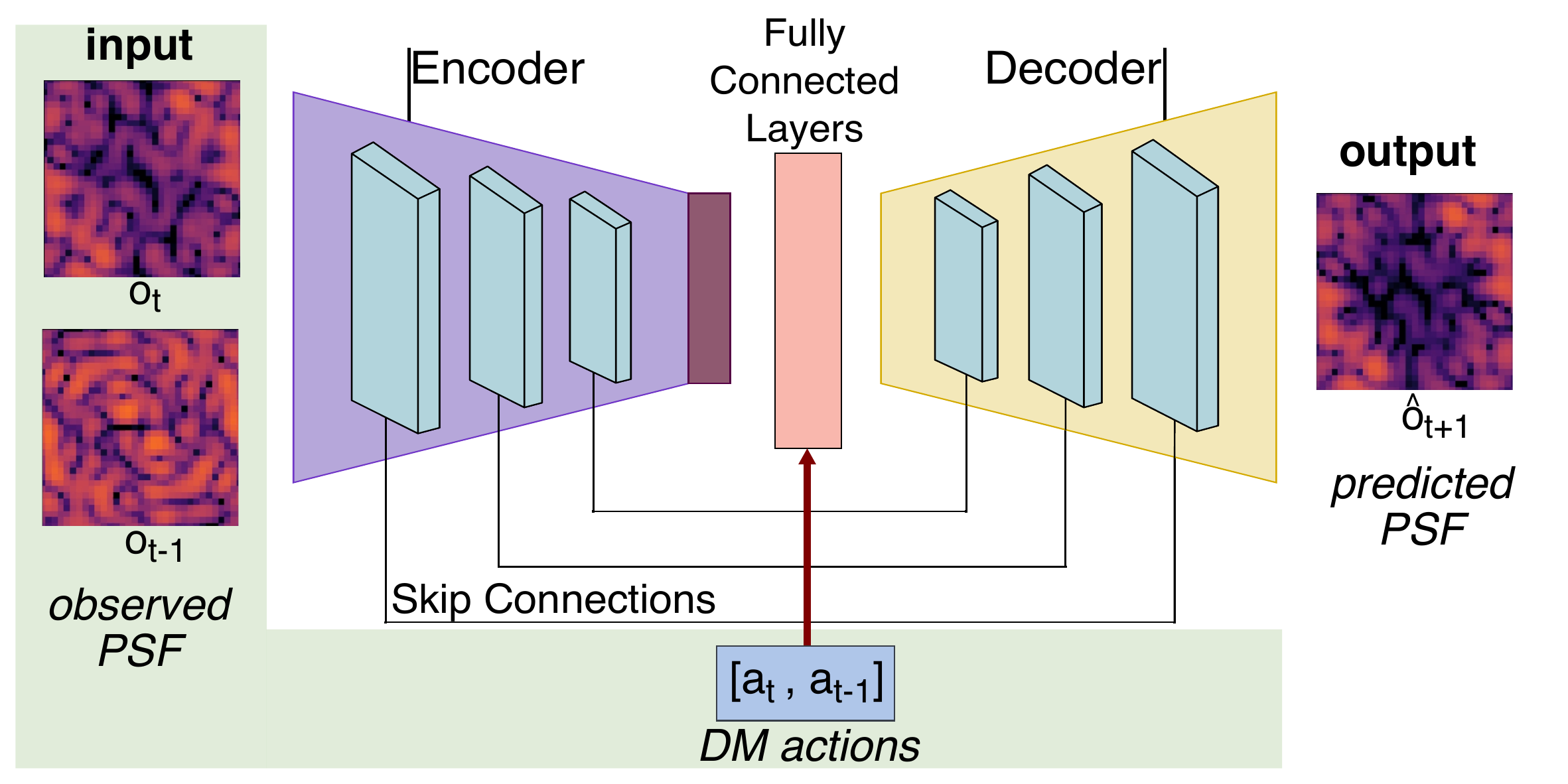} & \includegraphics[height=4.5cm]{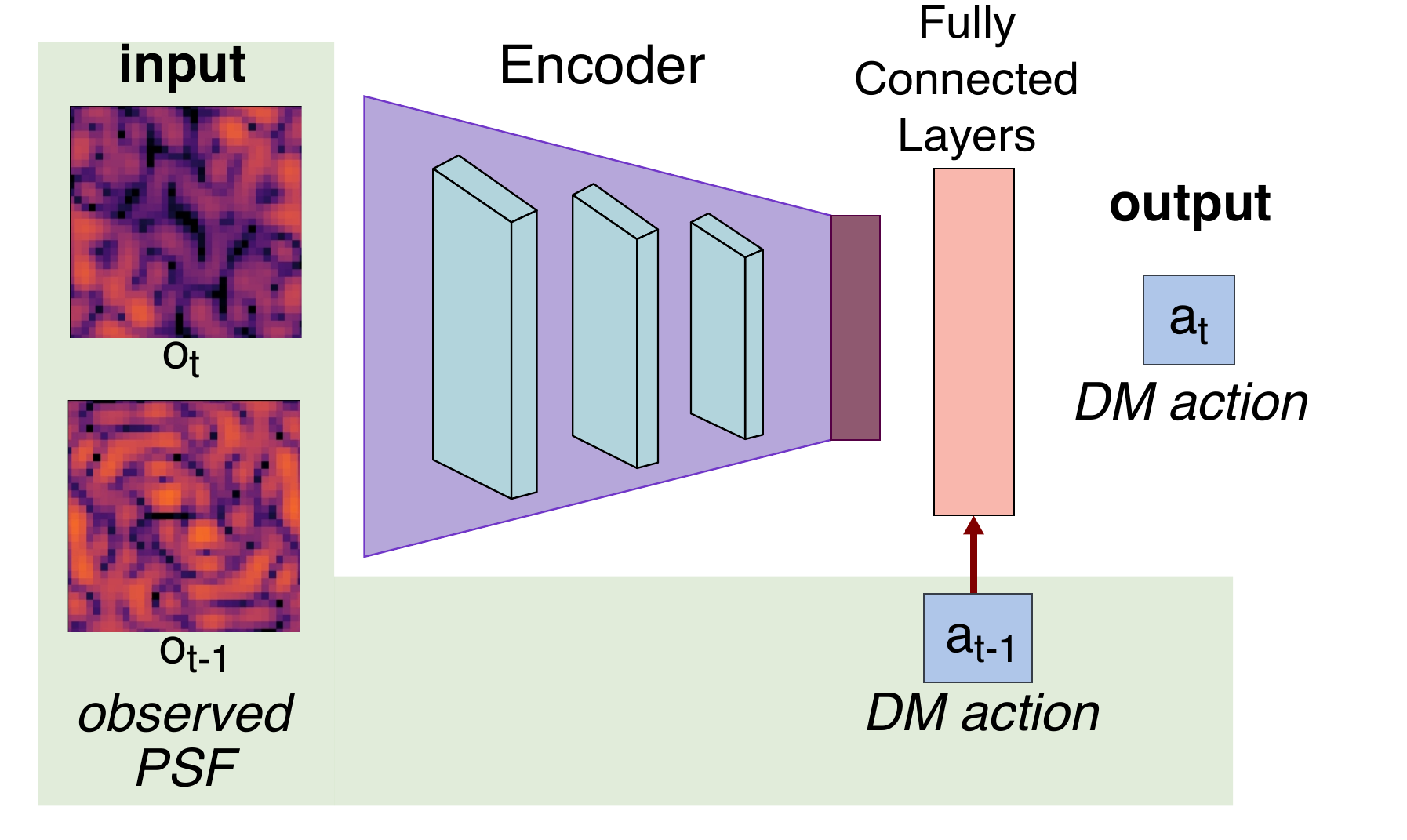}
   \end{tabular}
   \end{center}
   \caption[example] 
   { \label{fig:cnn_models} 
NN designs of the dynamics model (left) and policy model (right). The green highlights represent the inputs of the NN. The dynamics model is trained on focal plane images of the science camera ($\boldsymbol{o}_t$, $\boldsymbol{o}_{t-1}$) and the DM commands ($\boldsymbol{a}_t$, $\boldsymbol{a}_{t-1}$) in order to learn how to simulate the optical path and provide the next focal plane image ($\hat{\boldsymbol{o}}_{t+1}$). The policy model uses the science camera images ($\boldsymbol{o}_t$, $\boldsymbol{o}_{t-1}$) and the DM command ($\boldsymbol{a}_{t-1}$) to decide on the future DM input ($\boldsymbol{a}_t$). }
   \end{figure}

Similar to our previous work in Ref.~\citenum{2026Nousiainen}, we use an iterative process to control and optimize our algorithm. This process heavily depends on focal-plane images, though modified to suit our Neural Network (NN). We subtract the diffraction pattern from the focal plane image and flatten the resulting image, which is referred to as observations and denoted as $\boldsymbol{o}_t$. We first train a dynamics model $p(\boldsymbol{s}_t, \boldsymbol{a}_t)$ that predicts the next focal-plane image $(\hat{\boldsymbol{o}}_{t+1})$ based on prior observations $(\boldsymbol{o}_t, \boldsymbol{o}_{t-1})$ and DM actions $(\boldsymbol{a}_t, \boldsymbol{a}_{t-1})$. We set the reward function as
\begin{equation}
\label{eq:reward}
r(\boldsymbol{s}_t, \boldsymbol{a}_t) = - ||\hat{\boldsymbol{o}}_{t+1}||^2.
\end{equation}
We base this reward on the negative norm between the ideal PSF and the observed PSF. The dynamics model is then used to optimize the policy $\pi(\boldsymbol{a}_t|\boldsymbol{s}_t)$. These two models are illustrated in Fig.~\ref{fig:cnn_models}. 

The policy optimization occurs in three distinct steps. First, the policy runs the control loop for an episode of $T$ timesteps. Then, the dynamics model is optimized using a loss function \cite{2020Landman_Haffert}. Finally, the policy is optimized using the dynamics model. In each iteration, the focal plane images and DM commands at each timestep are stored. This data is then used to update the dynamics and policy models. Based on our results in Ref.~\citenum{2026Nousiainen}, we use 20 steps per episode to allow the algorithm to learn the actions. We use a CNN structure based on Ref.~\citenum{2022Nousiainen} while designing our architecture. We note that training VVC simulations take twice as long as trainignSVC simulations. 

\subsection{Simulation Description}
\label{sec:Simul}

To evaluate the performance of PO4NCPA, we use numerical simulations. We make use of the HCIPy \cite{por2018hcipy} python package to simulate the telescope, relevant optical components (coronagraph, deformable mirror (DM)) and atmospheric water vapour aberrations. We use the latest release of the ELT pupil by ESO where one of the spider arms is thicker (386 mm instead of 202 mm) due to the presence of the primary mirror crane. The DM is modelled to apply a command vector of Zernike modes. We use 54 Zernike modes for our numerical simulations, disregarding the piston mode. The decisions made on the Zernike mode selection are further discussed in Ref.~\citenum{2026Nousiainen}. The focal plane sampling is 8 pixels per $\lambda/D$, and the image is cropped to contain 5.5~$\lambda/D$. This produces a 33 x 33 pixel focal plane that covers the entire DM control radius for 54 Zernike modes. We select a frame rate of 10 Hz for dynamic NCPAs based on the recommendation of Ref.~\citenum{2024Orban_asymm} for WV seeing. We consider photon noise and background noise, but do not take AO residuals into consideration in this preliminary study. 

In order to ensure consistency across our numerical experiments, we use the same spatial NCPA spectrum for both static and dynamic NCPAs. As METIS conducts N-band observations at 11~\textmu m, it is primarily affected by WV effects \cite{2026Absil_SPIE}. The atmospheric parameters are based on the WV spectrum from Ref.~\citenum{2024Orban_asymm}. The WV phase screens are generated from a Kolmogorov spectrum with Fried parameter of 95~m and an outer scale of 500~m, to match the median PSD for WV seeing\cite{2024Orban_asymm}. This results in a median root mean squared error (RMSE) of 266~$\pm$~9~nm. For static NCPA simulations, a single phase screen is generated per episode for the policy to correct. For dynamic NCPA simulations, we propagate a single atmospheric layer at a wind speed of 10~m/s according to Taylor's frozen flow hypothesis. This corresponds to a coherence time of 2.9~s at 11~$\mu$m. In all simulations, we subtract the piston mode from the phase screens. All simulation parameters are listed in Tab.~\ref{table:simulator_parameters}. 

\begin{table}[ht]
    \centering
    \caption{ Simulation parameters}
    \label{table:simulator_parameters}
\begin{tabular}{ l l l} 
 \hline\hline
 \multicolumn{3}{c}{Telescope parameters} \\
 \hline
         Parameter  & Value  &  Units  \\
 \hline
 Telescope diameter  &  39.3  & m     \\
 Sampling frequency   &  10  & Hz        \\
  DM influence functions    &  ``Zernike''  & -  \\
 Number of modes     &  54   & -         \\
 Wavelength  & 11 & \textmu  m \\
 Star flux ($\textrm{mag}_N=1$) &   $3.67 \times 10^9$
   &   \#/frame          \\ 
 Background noise & $1.00 \times 10^8$  & \#/pix/frame  \\
 \hline
 \multicolumn{3}{c}{Water vapor seeing parameters} \\
 \hline
 Fried parameter ($r_0$)     &  95  & m @ 11~\textmu  m    \\
 Wind speed   &  10  & m/s    \\
 Outer scale ($L_0$)      & 500  &   m        \\ 
 Median wavefront RMSE   &  266 $\pm$ 9  & nm  \\
  \hline
 \multicolumn{3}{c}{PO4NCPA parameters} \\
 \hline
 Horizon limits $(h_{\text{min}},h_{\text{max}})$     &  (2,4)  & steps     \\
 CNN ensemble size        & 5   &   -         \\   
 Dynamics iterations / episode & 8 & steps\\
 Policy iterations / episode & 5 & steps \\
 Training minibatch size  & 64 &  - \\
  \hline 
\end{tabular}
\end{table}

With the given parameters, we conduct four experiments. We simulate two instrument setups, one with SVC and one with VVC. We test the performance of PO4NCPA in these two setups for static NCPAs and dynamic NCPAs. We analyze the performance of our algorithm through PSF-based (PSF contrast) and pupil-plane wavefront error-based (modal RMSE, mean modal error) metrics. The wavefront error is defined as 
\begin{equation}
\label{eq:err}
\mathrm{err}(x,t,i) = A_{\mathrm{WV}}(x,t,i) - A_{\mathrm{DM}}(x,t,i)
\end{equation}
where $A_{\mathrm{WV}}(x,y,i)$ is the 3D vector of WV coefficients for an experiment where $x$ is the number of episodes, $t$ is the number of timesteps and $i$ is the number of injected modes. In the same vein, $A_{\mathrm{DM}}(x,t,i)$ is the 3D vector of DM coefficients for the same experiment with the same parameters. To analyze our results, we consider an experiment with $x=X$ episodes, $t=T$ timesteps and any number of modes. We disregard the first few timesteps at each episode to allow the policy to converge. As such, we start our calculations from timestep $t_0$, up to timestep $T$. Using Eq.\ref{eq:err}, we calculate the mean and standard deviation per mode as
\begin{equation}
\label{eq:mean}
\mu_{\mathrm{err}}(i) = \frac{1}{XT} \sum_{(x,t) = (0,t_0)}^{(x,t)=(X,T)} \mathrm{err}(x,t,i),
\end{equation}
\begin{equation}
\label{eq:stdev}
\sigma_{\mathrm{err}}(i) = \sqrt{\frac{1}{XT}\sum_{(x,t) = (0,t_0)}^{(x,t)=(X,T)} \mathrm{err}(x,t,i) - \mu_{\mathrm{err}}(i)}.
\end{equation}
For the modal RMSE calculations, we use
\begin{equation}
\label{eq:rms_err}
\mathrm{RMS}_{\mathrm{err}}(i)= \sqrt{\frac{1}{XT} \sum_{(x,t) = (0,t_0)}^{(x,t)=(X,T)} \mathrm{err}^2(x,t,i)}.
\end{equation}
For our PSF contrast analysis, we calculate the radial contrast for the experiments. We compare the closed loop PO4NCPA correction to that of open loop (no correction), fitting error and a 1-step delay integrator (denoted as fitting $+$ delay error).

\section{RESULTS AND DISCUSSION}
\label{sec:results}

\subsection{Static NCPA}

After the training is complete, we ran $x=500$ episodes of $t=20$ timesteps and used the final policy to correct the aberrations. At each episode, we generated a randomly sampled atmospheric phase screen and a flat DM. We analyzed our results based on the residual modes and the PSF contrast.

\begin{figure} [t]
   \begin{center}
   \begin{tabular}{ll} 
   \includegraphics[height=4.2cm]{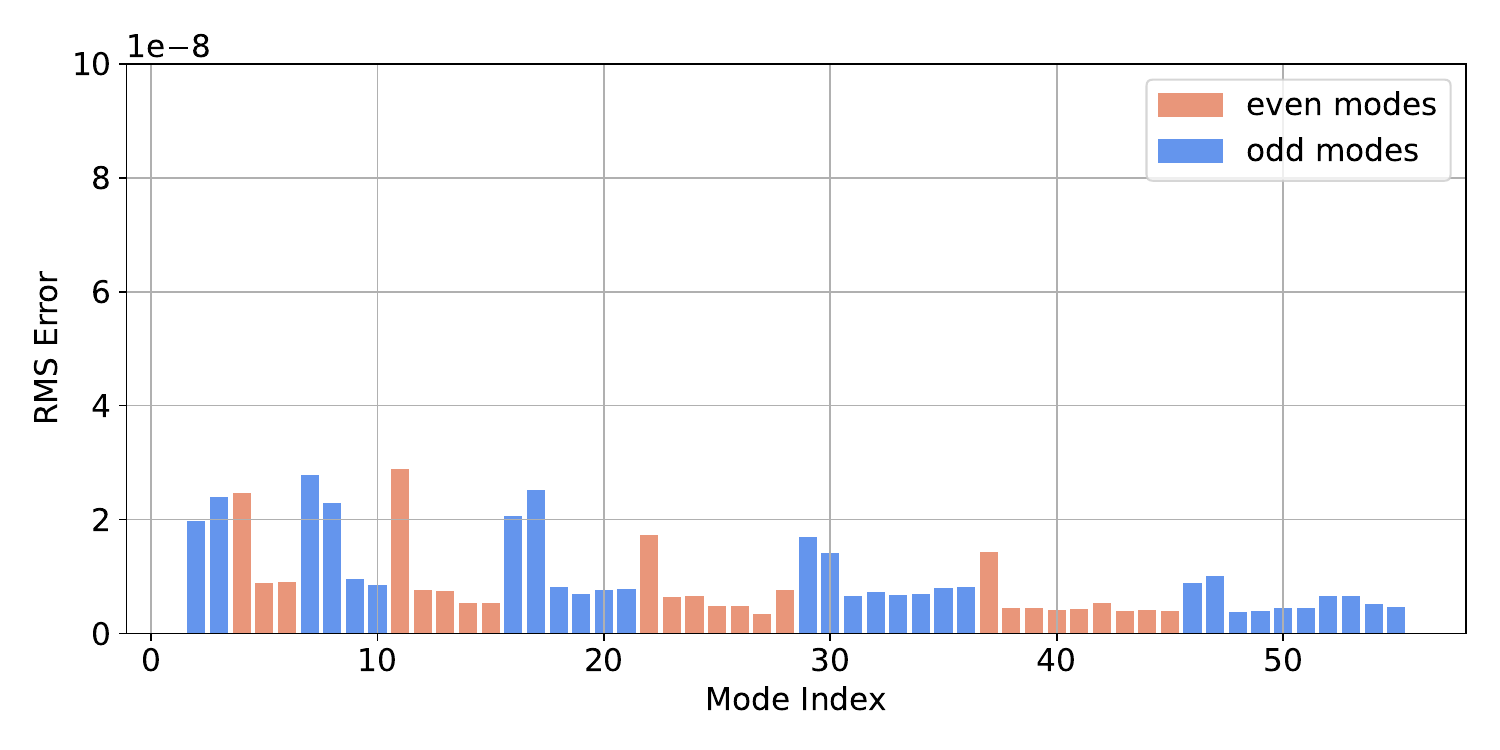} & \includegraphics[height=4.2cm]{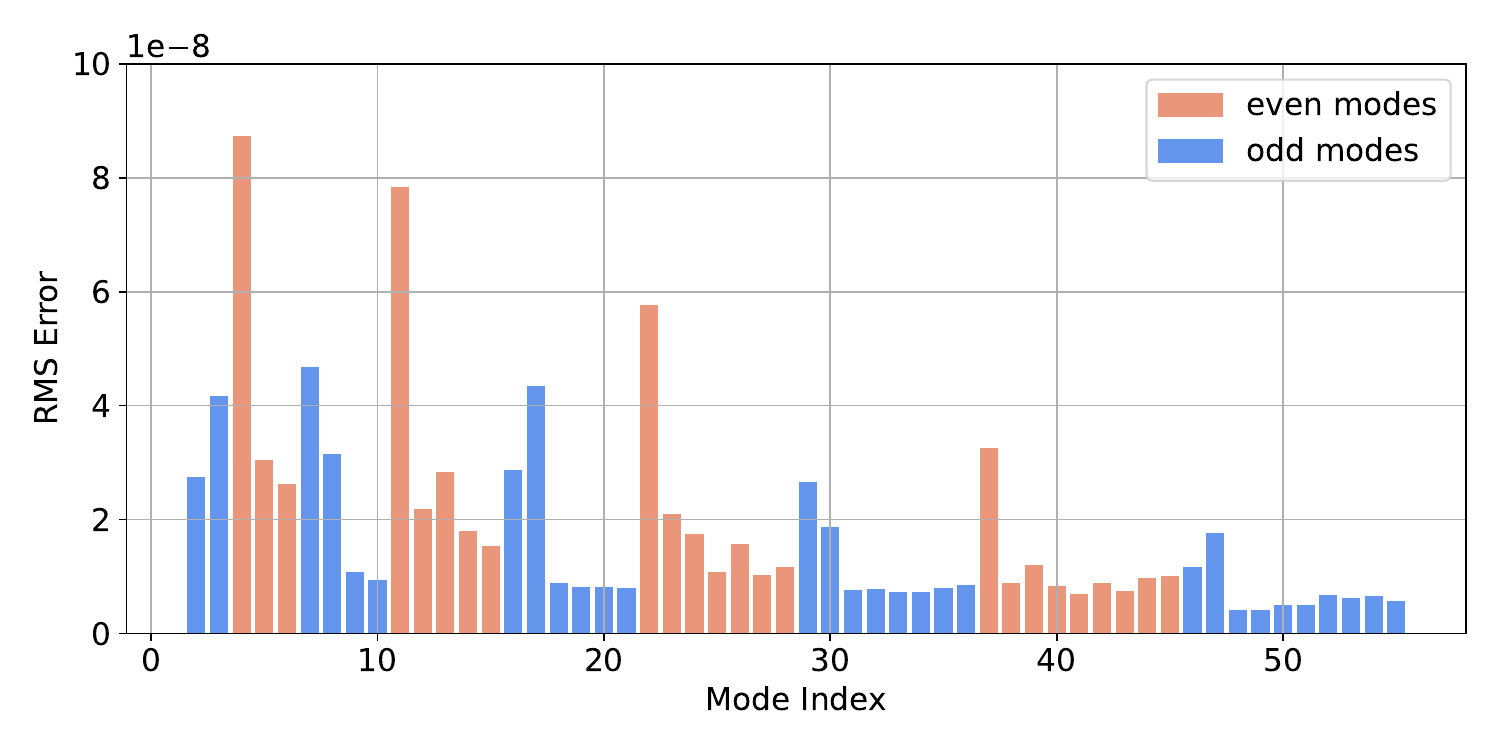}
   \end{tabular}
   \end{center}
   \caption[example] 
   { \label{fig:modes_static} 
RMS modal error for SVC (left) and VVC (right) with static NCPA. Orange bars represent the even radial order Zernike modes while the blue bars represent the odd radial order Zernike modes. The modal error is calculated across 500 episodes of 20 timesteps.}
   \end{figure}

To analyze our modal results, we used Eq.\ref{eq:mean}, Eq.\ref{eq:stdev} and Eq.\ref{eq:rms_err} where the episode number is $X=500$ with 17 timesteps per episode. We note that the first three timesteps of each episode are excluded in our analysis to allow the policy stabilize, thus the $T=17$ timesteps. Fig.~\ref{fig:modes_static} displays the RMS modal error plots for the two coronagraphs.

The median final RMSE after correction for the SVC is 74.9~$\pm$~33~nm. As mentioned in Sec.~\ref{sec:vortex}, phase retrieval with the SVC is ambiguous to pure radial even modes ($n$ even and $m=0$), corresponding to the first mode of each even radial order. This is  visible in Fig.~\ref{fig:modes_static} as pure radial even modes (first orange bar of each even radial order) show slightly higher RMS modal error values compared to the other even modes. For the VVC, the median final RMSE after correction is 131.2~$\pm$~49~nm. The VVC experiment has significantly higher RMSE compared to SVC, which can be largely attributed to the ambiguity on all even Zernike modes in the case of VVC. This ambiguity also leads to a degraded sensitivity to odd modes, as already noted in Ref.~\citenum{2022Quesnel}. In both cases, we observe that the first two modes of each odd radial order display larger errors. So far, we do not have a clear explanation of this result and will need to further analyze the policy behavior for possible clues. Finally, we emphasize that the policy algorithm aims to optimize the contrast rather than minimize the wavefornt error. Thus, these results are merely reflections of the behavior of the SVC and VVC for even modes.

\begin{figure} [ht]
   \begin{center}
   \begin{tabular}{c c} 
   \includegraphics[height=8cm]{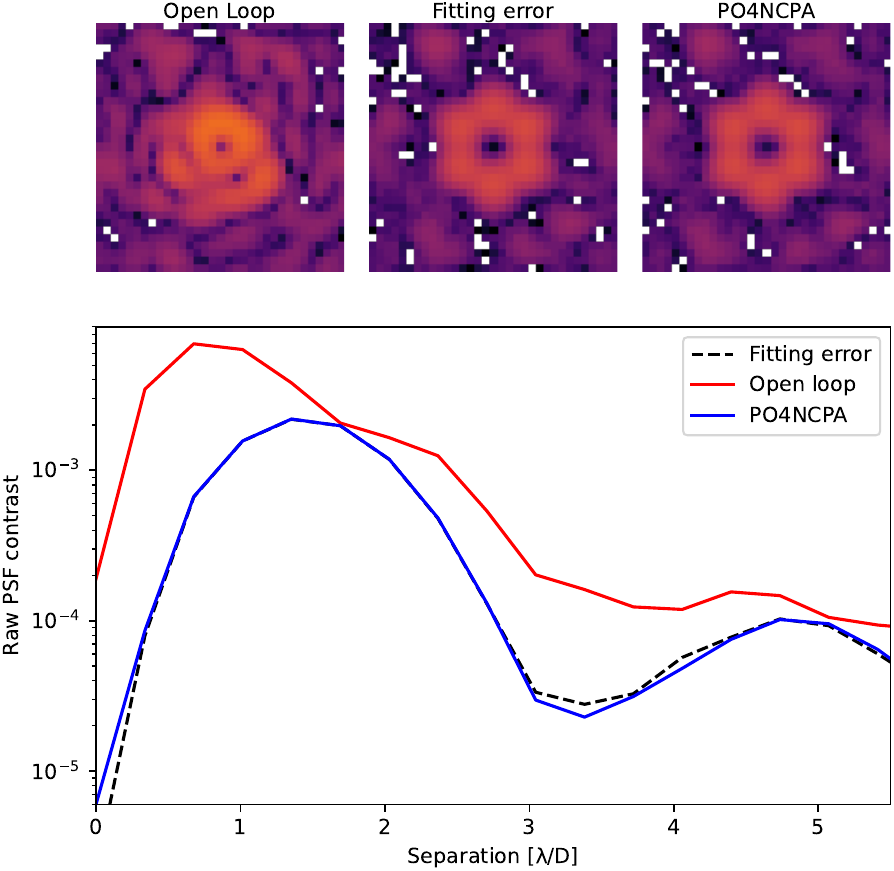} & \includegraphics[height=8cm]{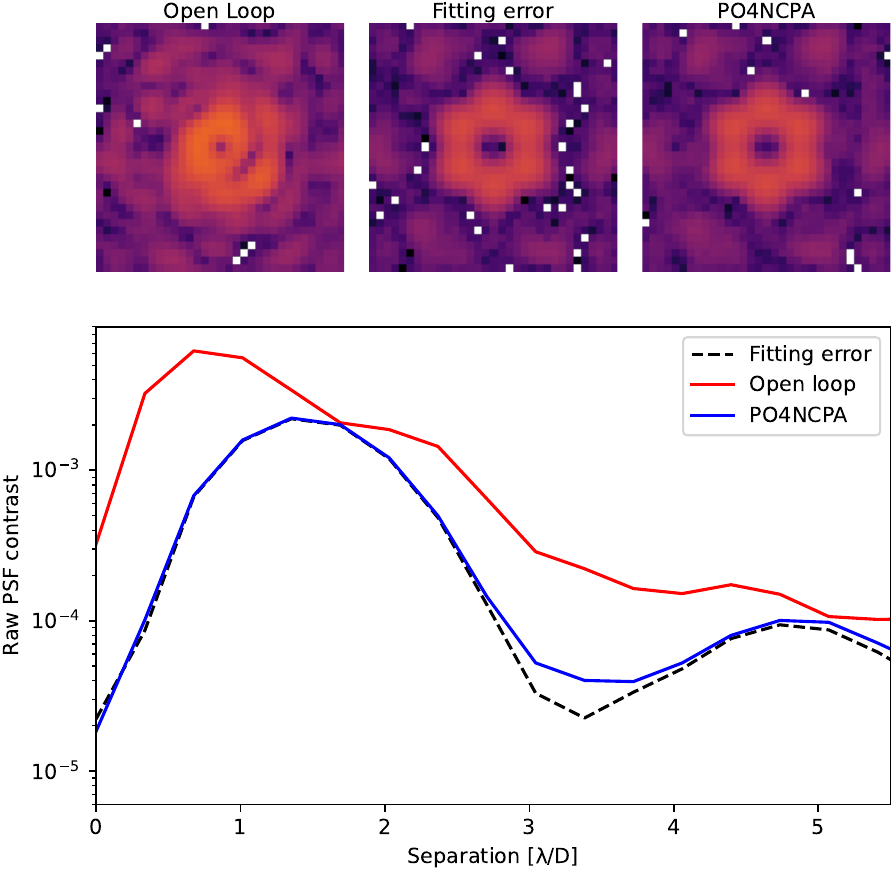} 
   \end{tabular}
   \end{center}
   \caption[example] 
   { \label{fig:psf_static} 
PSF sharpness and raw contrast for SVC (left) and VVC (right) with static NCPA. We plot the PSFs at the last timestep on top of the contrast curves. For the contrast curves, we take the average PSF across 17 timesteps (excluding the first 3 to allow the policy to converge to a result) and plot the radial average divided by the peak intensity.}
   \end{figure}

Fig.~\ref{fig:psf_static} illustrates the PSF images and plots the raw contrast for static NCPA experiments. In SVC experiments, open loop Strehl ratio is 98.85\% and fitting error Strehl ratio is 99.74\%. With PO4NCPA, the Strehl ratio closely follows fitting error at 99.57\%. Overall, PO4NCPA follows the fitting curve very closely, acquiring slightly better contrast between $2.8\lambda/\mathrm{D}-4.5\lambda/\mathrm{D}$. In VVC experiments, the Strehl ratio is 98.85\% for open loop while it is 99.81\% for the fitting error. In closed loop, PO4NCPA is able to bring the Strehl ratio to 99.27\%. While PO4NCPA performs very similar to the fitting overall, it is worse than fitting between $2.4\lambda/\mathrm{D}-3.8\lambda/\mathrm{D}$.

\subsection{Dynamic NCPA}

Similar to the static NCPA experiments, we used the final policy to run $X=10$ episodes of 5000 timesteps for the dynamic NCPA experiments. Generating and propagating an atmosphere for 5000 timesteps takes up quite a long time and has a large computational load. Thus, we made the decision to only run 10 individual episodes for our dynamic NCPA experiments. As we have done in the static NCPA experiments, we analyze our results through residual modes and PSF contrast.

\begin{figure} [ht]
   \begin{center}
   \begin{tabular}{ll} 
   \includegraphics[height=4.2cm]{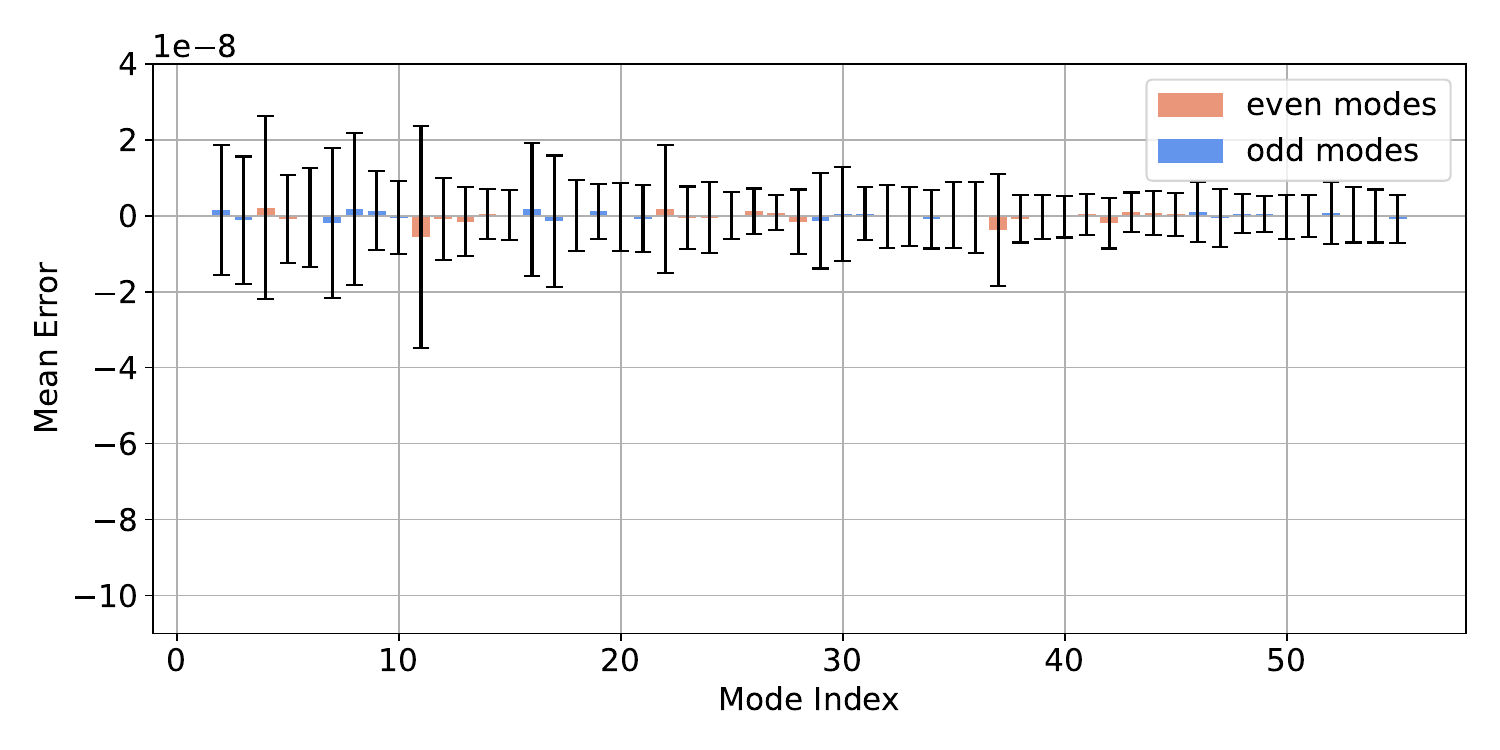} & \includegraphics[height=4.2cm]{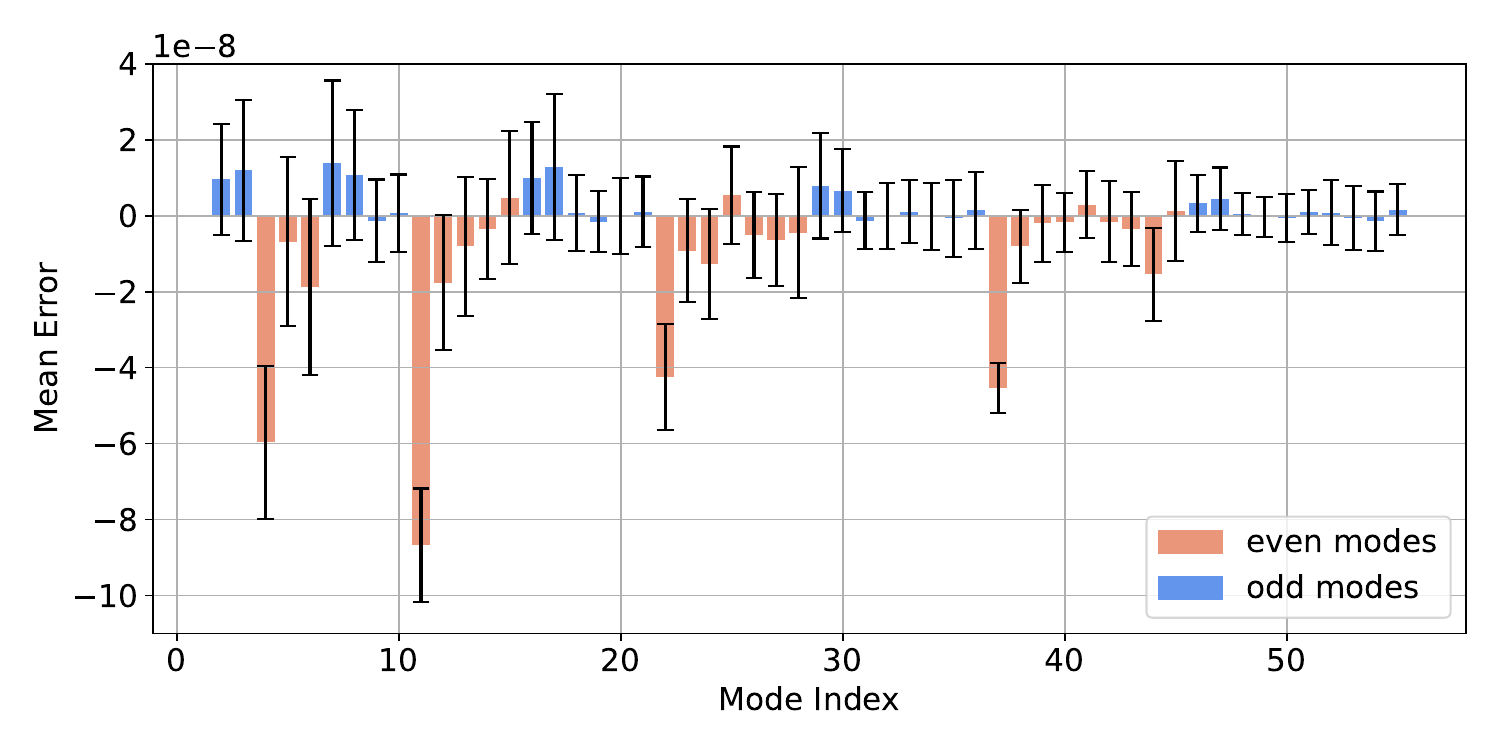}
   \end{tabular}
   \end{center}
   \caption[example] 
   { \label{fig:modes_dynamic} 
Mean modal error with standard deviation for SVC (left) and VVC (right) with dynamic NCPA. The orange bars represent the even radial order Zernike modes while the blue bars represent the odd radial order Zernike modes. Here, we take 4500 timesteps into account, disregarding the first 500 timesteps allowing the policy and the closed loop correction to converge.}
   \end{figure}

To analyze our modal results, we once again used Eq.\ref{eq:mean}, Eq.\ref{eq:stdev} and Eq.\ref{eq:rms_err} where the episode number is $X=10$ with $T=4500$ timesteps per episode. Here, too, we exclude the first 500 timesteps in each episode in order to ensure that the policy has converged, thus the $t=4500$ timesteps. Fig.~\ref{fig:modes_dynamic} illustrates the mean modal error for the SVC and VVC. The median RMSE after correction for the SVC is 67.6~$\pm$~22~nm, but fluctuates between a minimum of 42.8~nm and a maximum of 121.7~nm. RMS modal error plots (not shown here) display similar results to the static NCPA case where even modes remain low with the exception of the pure radial even modes. The mean modal error, illustrated in Fig.~\ref{fig:modes_dynamic}, calculates very small values for the SVC as it averages out over 4500 steps. For the VVC, the median RMSE after correction is 105.8~$\pm$~17~nm but fluctuates between 71.2~nm and 219.7~nm. Unlike the SVC, the VVC shows considerably higher mean error values for all even modes. Calculating the mean modal error in Fig.~\ref{fig:modes_dynamic}, we see that the pure radial even modes exceed the standard deviation. This could imply that either the policy is introducing a fixed diversity or it is insufficient in correcting certain modes. In follow-up work, we will investigate the actions of the policy in detail. 

\begin{figure} [ht]
   \begin{center}
   \begin{tabular}{c c} 
   \includegraphics[height=8cm]{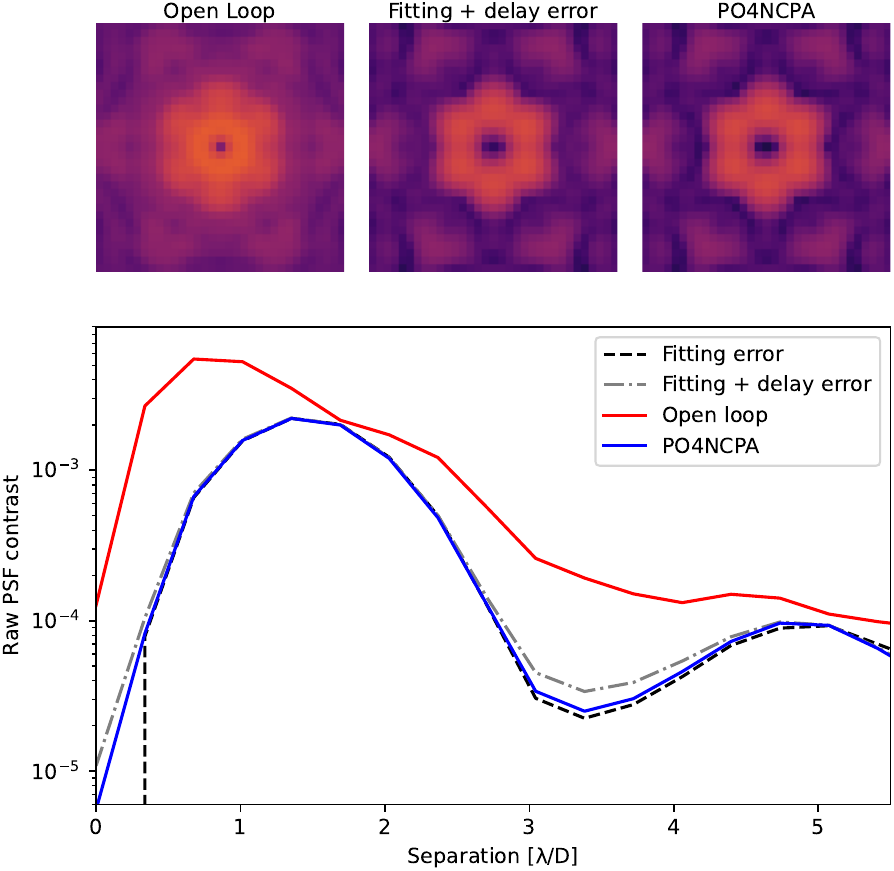} & \includegraphics[height=8cm]{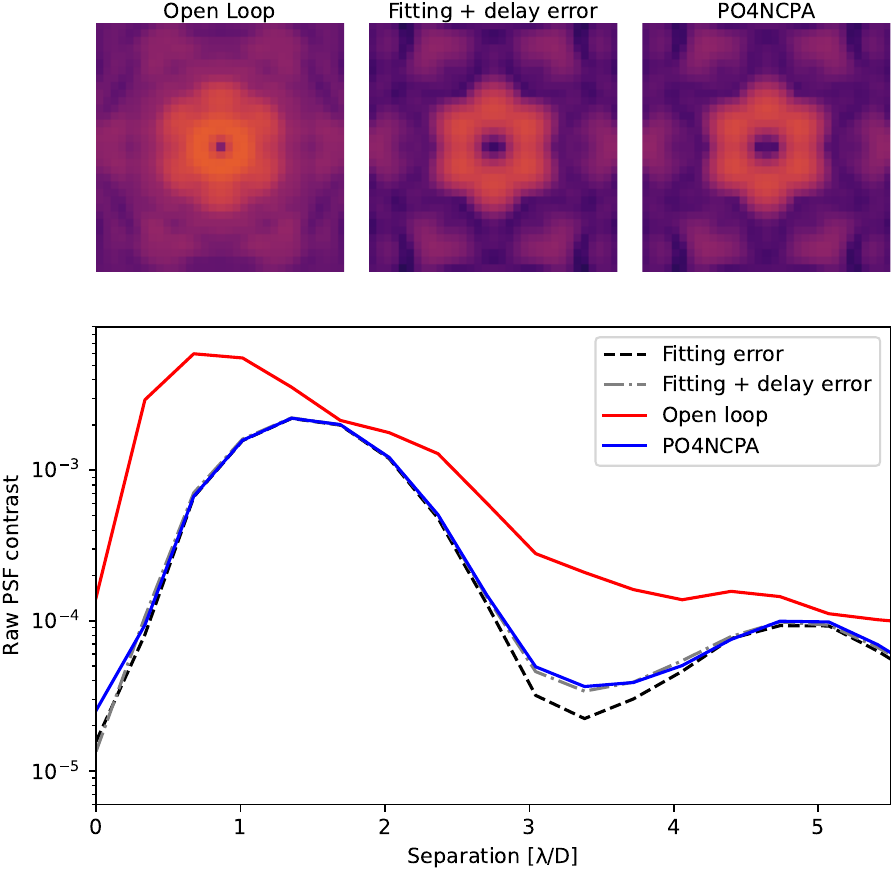} 
   \end{tabular}
   \end{center}
   \caption[example] 
   { \label{fig:psf_dynamic} 
PSF sharpness and raw contrast for SVC (left) and VVC (right) with dynamic NCPA. Here, we take the average PSF across 4500 timesteps (excluding the first 500 to avoid possible initial fluctuations in the correction) and plot the resulting PSF. For the raw PSF contrast, we plot the radial average divided by the peak intensity.}
   \end{figure}

Fig.~\ref{fig:psf_dynamic} plots the PSF images and the raw contrast for dynamic NCPA. SVC experiments show an open loop Strehl of 98.85\% and a fitting error Strehl of 99.74\%. When we take into account the delay error and the fitting error, the Strehl ratio is 99.63\%. According to Fig.~\ref{fig:psf_dynamic}, PO4NCPA follows the fitting error very closely across most of the FOV. For separation $2.8\lambda/\mathrm{D}-5\lambda/\mathrm{D}$, PO4NCPA is able to achieve better contrast than the fitting $+$ delay curve. This could point to possible predictive behavior on the policy's part. The VVC experiments have an open loop Strehl of 98.85\% and a fitting error Strehl of 99.81\%. The Strehl ratio for fitting error and delay error is 99.62\%. PO4NCPA is able to bring the Strehl ratio to 99.24\% in closed loop. It follows the delay $+$ fitting error curve very closely, but the fitting curve performs better overall. The difference between the two behaviours is probably linked to the fact that the SVC allows instantaneous phase retrieval by lifting the sign ambiguity on (most) even modes, while the VVC needs to rely on sequential phase diversity to lift this ambiguity (as in Ref.~\citenum{2026Nousiainen}), leading to a slower behaviour in the time domain.

\section{CONCLUSION}

We tested PO4NCPA, originally introduced in Ref.~\citenum{2026Nousiainen}, with SVC and VVC to observe its performance with different coronagraphs. We saw that PO4NCPA is able to acquire near-optimal control for systems with SVC and VVC. Compared to our previous experiments and the SVC experiments, we saw that training VVC takes considerably longer amount of time, up to several days depending on the amount of episodes that are trained. 

In both static and dynamic NCPA cases, PO4NCPA is able to achieve near-perfect Strehl. In all four experiments, we saw that PO4NCPA's correction follows very closely the correction of a 1-step delay integrator (denoted previously as fitting $+$ delay error). Considering the very high background noise injected into the simulations, we can say that PO4NCPA is quite robust in the presence of noisy observations. We also see possible predictive control in SVC experiments with dynamic NCPA. The modal analysis and the contrast difference between SVC and VVC experiments point to the SVC's ability to break the sign ambiguity for non-pure radial even modes (as predicted in Refs.~\citenum{2022Quesnel, 2024Orban_vortex}), which allows PO4NCPA to acquire higher contrast. 

Our immediate next steps are to perform a detailed analysis to characterize the policy behavior. Following this characterization, we will be implementing asymmetric masks instead of a circular Lyot stop in an attempt to lift the sign ambiguity in the VVC context, as foreseen for METIS\cite{2024Orban_asymm}. In order to get a more accurate representation of the ELT, we can add AO residuals, LWE and amplitude errors to our next numerical simulations. Combining these with the METIS asymmetric Lyot stop mask, we can perform numerical simulations that can closely represent how PO4NCPA would perform on the ELT/METIS instrument. A study of the performance of PO4NCPA on different stellar magnitudes expected to be observed by METIS is also an area of interest. A different approach to investigate will be to redesign the reward function in order to dig a dark hole. 

In addition to numerical simulations, laboratory and on-sky tests are natural to pursue in our future work. Considering the lengthy training time of PO4NCPA, on-sky training is not sustainable. Alternative training procedures need to be pursued, such as a two step procedure of initial simulation training followed by fine tuning on the instrument during daytime. Comparing the performance of PO4NCPA when trained on simulation compared to when trained on-sky is also an interesting experiment to test the adaptability of the policy. On this front, we aim to work on the GHOST bench of ESO and the SCExAO instrument of the Subaru telescope in the future.

\newpage

\acknowledgments 
This research is funded by Le Fonds de la Recherche Scientifique (FNRS). The attendance and travel of İremsu Taşkın to SPIE Astronomical Telescopes and Instrumentation 2026 conference is kindly funded by SPIE Student Conference Support. İ.T.\ is a FRIA Grantee of the Fonds de la Recherche Scientifique – FNRS. O.A.\ is a Research Director of the Fonds de la Recherche Scientifique – FNRS.

\bibliography{report} 
\bibliographystyle{spiebib} 

\end{document}